\documentclass[11pt, a4]{article}

\usepackage{graphicx, hyperref}
\usepackage{amsfonts, amsmath,amssymb, mathrsfs, wasysym}
\usepackage[english]{babel}
\usepackage{datetime}
\usepackage[normalem]{ulem}
\usepackage{color,slashed}

\ifx\pdfoutput\undefined
\usepackage[dvips,bookmarks]{hyperref}	
\else
\usepackage{hyperref}	
\fi
\hypersetup{colorlinks,bookmarksopen,bookmarksnumbered,citecolor=rossos,
linkcolor=rossos,pdfstartview=FitH,urlcolor=rossos}

\def\circa#1{\,\raise.3ex\hbox{$#1$\kern-.75em\lower1ex\hbox{$\sim$}}\,}

\numberwithin{equation}{section} 
\usepackage[table]{xcolor}

\definecolor{grigino}{cmyk}{0,0,0,0.2}
\definecolor{mentuccia}{cmyk}{0.4,0,0.3,0.1}
\definecolor{arancino}{cmyk}{0,0.1,0.4,0}
\definecolor{menta}{cmyk}{0.7,0,0.5,0.3}
\definecolor{grigios}{cmyk}{0,0,0,0.5}
\definecolor{bianco}{cmyk}{0,0,0,0}
\definecolor{arancio}{cmyk}{0,0.2,0.6,0}
\definecolor{grigio}{cmyk}{0,0,0,0.1}
\definecolor{rosa}{cmyk}{0,0.1,0.1,0.02}
\definecolor{rosino}{cmyk}{0,0.05,0.05,0.02}
\definecolor{rosas}{cmyk}{0,0.3,0.25,0.05}
\definecolor{celeste}{cmyk}{0.1,0,0,0.02}
\definecolor{giallino}{cmyk}{0,0,0.4,0.02}
\definecolor{rosso}{cmyk}{0,1,1,0.4}
\definecolor{rossos}{cmyk}{0,1,1,0.55}
\definecolor{rossoc}{cmyk}{0,1,1,0.2}
\definecolor{blu}{cmyk}{1,1,0,0.3}
\definecolor{blus}{cmyk}{1,1,0,0.5}
\definecolor{bluc}{cmyk}{1,1,0,0.1}
\definecolor{blucc}{cmyk}{0.7,0.5,0,0}
\definecolor{viola0}{cmyk}{0,0.4,0,0.04}
\definecolor{viola}{cmyk}{0,0.5,0,0.05}
\definecolor{viola2}{cmyk}{0,1,0.2,0.6}
\definecolor{verde}{cmyk}{0.92,0,0.59,0.25}
\definecolor{verdec}{cmyk}{0.92,0,0.59,0.15}
\definecolor{verdecc}{cmyk}{0.42,0,0.8,0.05}
\definecolor{verdes}{cmyk}{0.92,0,0.59,0.4}
\definecolor{verdino}{cmyk}{0.12,0,0.3,0.02}
\definecolor{giallo}{cmyk}{0,0,1,0}
\definecolor{gialloverde}{cmyk}{0.44,0,0.74,0}

\newcommand{\be}{\begin{equation}}
\newcommand{\ee}{\end{equation}}
\newcommand{\bea}{\begin{eqnarray}}
\newcommand{\eea}{\end{eqnarray}}

\newcommand{\newc}{\newcommand}

\newc{\gsim}{\lower.7ex\hbox{$\;\stackrel{\textstyle>}{\sim}\;$}}
\newc{\lsim}{\lower.7ex\hbox{$\;\stackrel{\textstyle<}{\sim}\;$}}

\numberwithin{equation}{section}

\def\fun#1#2{\lower3.6pt\vbox{\baselineskip0pt\lineskip.9pt
  \ialign{$\mathsurround=0pt#1\hfil##\hfil$\crcr#2\crcr\sim\crcr}}}
\def\simgt{\mathrel{\lower0.6ex\hbox{$\buildrel {\textstyle >}
 \over {\scriptstyle \sim}$}}}
\def\simlt{\mathrel{\lower0.6ex\hbox{$\buildrel {\textstyle <}
 \over {\scriptstyle \sim}$}}}
\def\mK{{\mathcal K}}
\def\bea{\begin{eqnarray}}
\def\eea{\end{eqnarray}}
\def\be{\begin{equation}}
\def\ee{\end{equation}}
\def\tr{{\rm tr}\,}
\def\mM{{\mathcal M}}

\input epsf

\def\be{\begin{equation}}
\def\ee{\end{equation}}
\def\ba{\begin{eqnarray}}
\def\ea{\end{eqnarray}}

\def\a{\alpha}

\begin{document}

\date{\mbox{}}

\title{
\vspace{-2.0cm}
%
\vspace{2.0cm}
{\bf \huge Strong interactions and exact solutions\\ in non-linear massive gravity
}
 \\[8mm]
}

\author{
Kazuya Koyama,\, Gustavo Niz,\, Gianmassimo Tasinato
\\[8mm]
\normalsize\it
Institute of Cosmology \& Gravitation, University of Portsmouth,\\
\normalsize\it Dennis Sciama Building, Portsmouth, PO1 3FX, United Kingdom
}

\maketitle

\setcounter{page}{0}
\thispagestyle{empty}

\begin{abstract}
\noindent
We investigate strong coupling effects in a covariant massive gravity model, which is a candidate for a ghost-free non-linear completion of Fierz-Pauli. We analyse the conditions to recover general relativity via the Vainshtein mechanism in the weak field limit, and find three main cases depending on the choice of parameters. In the first case, the potential is such that all non-linearities disappear and the vDVZ discontinuity cannot be avoided. In the second case, 
the Vainshtein mechanism allows
to recover General Relativity within a macroscopic radius from a source.
In the last case, the strong coupling of the scalar graviton completely shields the massless graviton,  and weakens gravity when approaching the source. In the second part of the paper, we explore new exact vacuum solutions, that asymptote to de Sitter or anti de Sitter space depending on the choice of parameters. The 
curvature of the space 
is proportional to the mass of the graviton, thus providing a cosmological background
which may explain the present day acceleration in terms of the graviton mass. Moreover, by expressing  the potential for non-linear massive gravity in a convenient form, we also suggest possible connections with a higher dimensional framework.
\end{abstract}


\def\eq#1{eq.~(\ref{#1})}

\smallskip

\section{Introduction}
\label{Introduction}

Attempts to build a theory of massive gravity
 date back to the work by Fierz and Pauli (FP) in 1939 \cite{Fierz:1939ix}. They considered a mass term for linear gravitational perturbations, which is uniquely determined by requiring the absence of ghost degrees of freedom. The mass term breaks the gauge invariance of General Relativity (GR), leading to a graviton with five degrees of freedom instead of the two found in GR. There have been intensive studies in to what happens beyond the linearised theory of FP. In 1972, Boulware and Deser found a scalar ghost mode at the non-linear level, the so called sixth degree of freedom in the FP theory \cite{Boulware:1973my}. This issue has been re-examined using an effective field theory approach \cite{ArkaniHamed:2002sp}, where gauge invariance is restored by introducing St\"uckelberg fields. In this language, the  St\"uckelberg fields acquire non-linear interactions containing more than two time derivatives, signalling the existence of a ghost. In order to construct a consistent theory, non-linear terms should be added to the FP model, which are tuned so that they remove the ghost order by order in perturbation theory.

Interestingly, this approach sheds light on another famous problem with FP massive gravity; due to contributions of the scalar degree of freedom, solutions in the FP model do not continuously connect to solutions in GR, even in the limit of zero graviton mass. This is known as the van Dam, Veltman, and Zakharov (vDVZ) discontinuity \cite{vanDam:1970vg, Zakharov:1970cc}. Observations such as light bending in the solar system would exclude the FP theory, no matter how small
the graviton mass is. In 1972, Vainshtein \cite{Vainshtein:1972sx} proposed a mechanism to avoid this conclusion; in the small mass limit, the scalar degree of freedom becomes strongly coupled and the linearised FP theory is no longer reliable. In this regime, higher order interactions, which are introduced to remove the ghost degree of freedom, should shield the scalar interaction and recover GR on sufficiently small scales.

\smallskip
Until recently, it was thought to be impossible to construct a ghost-free theory for massive gravity that is compatible with current observations \cite{Creminelli:2005qk, Deffayet:2005ys}. A breakthrough came with a 5D braneworld model known as Dvali-Gabadadze-Porrati (DGP) model \cite{Dvali:2000hr}.
 In this model there appears a continuous tower of massive gravitons
 from a four dimensional perspective, and GR can be recovered for a given range of scales, due to strong coupling interactions \cite{Luty:2003vm, Nicolis:2004qq}.
In this paper, we explore the consequences of a promising new development along these lines that seem able to
provide a consistent theory of massive gravity directly in four dimensions. 
 In order to avoid the presence of a ghost, interactions have to be chosen
in such a way that the equations of motion for the scalar degrees of freedom contain no more than two time derivatives. Recently, it was shown that there is a finite number of derivative interactions that give rise to second order differential equations. These are dubbed Galileon terms because of a symmetry under a constant shift of the scalar field derivative \cite{Nicolis:2008in}. Therefore, one expects that any consistent non-linear completion of FP contains these Galileon terms in the limit in which the scalar mode decouples from the tensor modes, the so-called {\it decoupling limit}. This turns out to be a powerful criterion for building higher order interactions with the desired properties. Indeed, following this route,  de Rham and Gabadadze constructed a family of ghost-free extensions to the FP theory, which reduce to the Galileon terms in the decoupling limit \cite{deRham:2010ik}.

\smallskip

In this work, we investigate the consequences of strong coupling effects in this theory. 
We first re-express the potential for the most general version of non-linear
massive gravity, as developed by de Rham, Gabadadze and collaborators, in a particularly compact, easy-to-handle form. Among other things, this way of writing the potential suggests intriguing relations with a higher dimensional set-up, which might offer new perspectives  for analysing this theory. Moreover, we show that, for certain parameter choices, the potential for massive gravity coincides with the action describing non-perturbative brane objects, independently supporting connections with a higher dimensional framework.

Armed with these tools, we then focus on the Vainshtein mechanism for this potential. 
We show that this theory is able to reproduce the behaviour of linearised solutions in 
General Relativity below the Vainshtein radius, but only in a specific region of parameter
space. This result provides stringent constraints on non-linear massive gravity.
Moreover, we are able to physically re-interpret these findings in the decoupling limit 
in terms of an effective theory with Galileon interactions. We show that the condition to successfully implement the Vainshtein mechanism is associated with the sign of a direct coupling between the massless graviton and the scalar degrees of freedom.

We also present new exact solutions in the vacuum that asymptote de Sitter or anti 
de Sitter space depending on the choice of the parameters. 
Asymptotically de Sitter configurations can be expressed in an explicit time-dependent form. These solution may provide an interesting background for the observed Universe
where the rate of the accelerated expansion of the Universe is set by the graviton mass.
A small graviton mass, as required by the solar system constraints on deviations from standard General Relativity, is then in agreement with the observed value of the cosmological constant. On the other hand, asymptotically anti de Sitter configurations may have interesting  applications to the AdS/CFT correspondence.

\smallskip

The paper is organized as follows: in Section \ref{covsect}, we discuss how to construct 
a non-linear potential for massive gravity and point out new connections with a higher dimensional set-up. In Section \ref{sect-vain}, we show how linearised Einstein's gravity is
recovered within a certain macroscopic radius from a mass source, via the Vainshtein mechanism. Only a subset of parameter space presents a successful Vainshtein effect in the weak field limit. For a better understanding of the theory and in particular of the ghost mode, it is important to find analytic non-linear solutions; we present new exact solutions in Section \ref{sec-exact}. Finally, we conclude in Section \ref{sect-discus}, leaving technical details of the calculations developed 
in the main text to the Appendixes.

\section{Covariant non-linear massive gravity}\label{covsect}

We start by introducing the covariant Fierz-Pauli mass term in four-dimensional spacetime
\be\label{lagFP}
\mathcal{L}_{FP}=m^2\sqrt{-g}\;{\cal U}^{(2)},\qquad\qquad {\cal U}^{(2)}=\left(H_{\mu\nu}H^{\mu\nu}-H^2\right),
\ee
where the tensor $H_{\mu \nu}$ is a covariantisation of the metric perturbations, namely
\be\label{H}
g_{\mu \nu} = \eta_{\mu \nu} +h_{\mu\nu}\,\equiv\,H_{\mu \nu}+
\partial_\mu \phi^\alpha \partial_\nu \phi^\beta \eta_{\alpha \beta}.
\ee
The St\"uckelberg fields $\phi^\alpha\,=\,\left(x^\alpha-\pi^\alpha\right)$
are introduced to restore  reparametrisation invariance, hence
transforming as scalars \cite{ArkaniHamed:2002sp}. The internal metric $\eta_{\alpha \beta}$ corresponds to a  non-dynamical reference
 metric,  usually assumed to be Minkowski space-time.
  Therefore, around flat space,  we can rewrite $H_{\mu \nu}$ as
\bea\label{defhmn}
H_{\mu \nu}&=&h_{\mu \nu}+\eta_{\beta \nu}\partial_\mu \pi^\beta+\eta_{\alpha \mu}\partial_\nu
\pi^\alpha-
\eta_{\alpha \beta} \partial_\mu \pi^\alpha \partial_\nu \pi^\beta, \nonumber\\
&\equiv& h_{\mu \nu}-{\mathcal Q}_{\mu\nu}.
\eea
From now on, indices are raised/lowered with the dynamical metric $g_{\mu\nu}$,
 unless otherwise
stated.
For example,
$H^\mu_{\,\,\nu}\,=\,g^{\mu \rho} H_{\rho \nu}$.
 Moreover, the Lagrangian (\ref{lagFP}) is invariant under coordinate transformations $x^{\mu} \to x^{\mu} + \xi^{\mu}$, provided
$\pi^{\mu}$ transforms as
\be\label{pitransf}
\pi^{\mu} \to \pi^{\mu} + \xi^{\mu}.
\ee
The  scalar component $\pi$ of the St\"uckelberg field
can be extracted from the relation
 $\pi^\mu=\eta^{\mu\nu}\partial_\nu\pi/\Lambda_3$, with $\Lambda_3^3=m^2M_{pl}$
  (the meaning of the scale $\Lambda_3$ will be explained in the following).
  The  dynamics
   of $\pi$ are 
  the
origin of the two problems discussed in the introduction: the DB ghost excitation and the vDVZ discontinuity.
  With respect to the first problem, as noticed
  by Fierz and Pauli, one can remove the ghost excitation, to linear order in perturbations, by choosing the
  quadratic
  structure $h_{\mu\nu}h^{\mu\nu}-h^2$.
   When expressed in
    the St\"uckelberg field language, by means
   of the scalar-graviton $\pi$, higher derivative
   terms  in the action
    are arranged in a such way to form  a total derivative,
    leading to second order
    equations of motion. However, when going beyond linear order, the equation of
    $\pi$ acquires  higher time derivatives, signalling the presence
     of a ghost mode \cite{ArkaniHamed:2002sp}. Remarkably,
     de Rham and Gabadadze  were able to
     construct a potential, tuned
     at each order in perturbations,
      to give a total derivative for the dangerous terms, leading to
     equations of motion that are at most second order in time
     derivatives \cite{deRham:2010ik}.

     \smallskip

     We now review
     their construction, introducing alternative ways to express the potential, which 
       provide
       a  new connection
      with a five dimensional point of view.
       In terms of the
 helicity zero mode, corresponding to the field $\pi$, we can write
 the tensor $H_{\mu \nu}$ of Eq.~ (\ref{defhmn}) as
\be
H_{\mu \nu} = h_{\mu \nu} + \frac{2}{M_{pl} m^2} \Pi_{\mu \nu}
- \frac{1}{M_{pl}^2 m^4} \Pi^2_{\mu \nu},
\label{Hscalar}
\ee
where $\Pi_{\mu \nu} = \partial_{\mu} \partial_{\nu} \pi$ and
$\Pi^2_{\mu \nu} = \Pi_{\mu \alpha} \Pi^{\alpha}_{\nu}$. At a given order $n$
in perturbations, the idea is to add terms of the form
\be\label{Un}
m^2\sqrt{-g}\;{\cal U}^{(n)}\;=\;
m^2\sqrt{-g}\;\sum_{i=0}^n c^n_i (H^{n-i}_{\mu\nu})(H^{i}),
\ee
to the FP action (\ref{lagFP}), and to choose the coefficients $c_i^n$
in order to get a total derivative for the leading contributions of the scalar mode, namely $(\Pi_{\mu\nu})^n$. The key finding of \cite{deRham:2010ik} is that these total derivatives are unique at each order, and that the series stops at quintic order in perturbations. 
 Let us describe in more detail the structure and origin of these terms. 
Following the notation of \cite{deRham:2010ik}, the total derivatives are given by
\be
{\cal L}^{(n)}_{der}=-\sum_{m=1}^{n}(-1)^m\frac{(n-1)!}{(n-m)!}(\tr \Pi^m_{\mu\nu}){\cal L}^{(n-m)}_{der},
\ee
with ${\cal L}^{(0)}_{der}= 1$ and
\bea
{\cal L}^{(1)}_{der}&=& \tr \Pi_{\mu\nu}, \nonumber\\
{\cal L}^{(2)}_{der}&=&
 (\tr \Pi_{\mu\nu})^2 - \tr \Pi_{\mu\nu}^2 \;, \nonumber \\
{\cal L}^{(3)}_{der}&=&
(\tr \Pi_{\mu\nu})^3 - 3 (\tr \Pi_{\mu\nu})(\tr \Pi_{\mu\nu}^2) + 2 \tr \Pi_{\mu\nu}^3 \;, \nonumber \\
{\cal L}^{(4)}_{der}&=&  (\tr \Pi_{\mu\nu})^4 - 6 (\tr \Pi_{\mu\nu})^2 (\tr \Pi_{\mu\nu}^2) \nonumber
+ 8 (\tr \Pi_{\mu\nu})(\tr \Pi_{\mu\nu}^3) + 3 (\tr \Pi_{\mu\nu}^2)^2 - 6 \tr \Pi_{\mu\nu}^4 \;.
\label{totalderiv}
\eea
and ${\cal L}^{(n)}_{der}$ vanishes for $n>4$. 
These expressions are related to a matrix determinant
(see also Note Added at the end of this paper).
To see this, consider a generic squared real
 matrix $A$, and a complex number $z$. Then,  the following formula holds
\be\label{detA}
\det\left( {\mathbb I}+ z A\right)=1+\sum_{n=1}^\infty z^i \det_n(A)
\ee
where $\det_{n} (A)$ can be written in terms  of traces  as
\bea
\label{dets}
\det_{1} (A) &=& \tr A, \nonumber\\
\det_{2} (A) &=& \frac{1}{2} \Big( (\tr A)^2 - \tr A^2 \Big), \nonumber \\
\det_{3} (A) &=& \frac{1}{6} \Big( (\tr A)^3 - 3 (\tr A)(\tr A^2) + 2 \tr A^3 \Big), \nonumber \\
\det_{4} (A) &=& \frac{1}{24} \Big( (\tr A)^4 - 6 (\tr A)^2 (\tr A^2)
+ 8 (\tr A)(\tr A^3) + 3 (\tr A^2)^2 - 6 \tr A^4 \Big).
\eea
Moreover,  all terms $\det_{n} (A) $ with $n>4$ vanish for a $4\times 4$ matrix. 
Therefore, for the choice $A_\mu^\nu=\Pi_\mu^\nu$ we get the simple relation $\mathcal{L}^{(n)}_{der}=n!\det_n(\Pi)$, and the series indeed stops at $n=4$.
  If one chooses a sum of determinants of the form
\be
\sum_{i=1}^4\det(\mathbb{I}+z_i\Pi)-4,
\ee
one can generate each $\det_n(\Pi)$ term with a separate coefficient $\beta_n$, provided a solution to $\sum_{i=1}^{4}z_i^n=\beta_n$ exists, which is guaranteed by the Newton identities. Then, the Lagrangian for the helicity zero mode (that
is, neglecting
for the moment the contributions of  tensor modes, and of the vector components
 of   the St\"uckelberg field) is
\be\label{lagpi}
\mathcal{L}_{\pi}=\sum_{n=1}^4\beta_n\det_n(\Pi).
\ee

\smallskip
We now briefly turn away from the present discussion, and show
 an interesting way to construct the Lagrangian (\ref{lagpi})
 from a higher dimensional point of view. Consider a
 five dimensional  Minkowski spacetime, and embed
 on it a test 3-brane  (i.e. we do not include
   back-reaction from brane dynamics). Under this assumption,  the
   five dimensional Riemann tensor vanishes;
   using the Gauss equation, the intrinsic curvature on the brane is related to the extrinsic curvature as
($R^{\alpha}_{\;\; \beta \gamma \delta}$ is
constructed in terms of  brane induced metric)
\be
R^{\alpha}_{\;\; \beta \gamma \delta} = K^{\alpha}_{\gamma} K_{\beta \delta} - K^{\alpha}_{\delta} K_{\beta \gamma}, \;\; R_{\mu \nu} = K K_{\mu \nu} - K^{\alpha}_{\mu} K_{\nu \alpha}, \;\; R = K^2 - K_{\mu \nu} K^{\mu \nu}.
\label{gauss}
\ee
We then consider a four dimensional  Lagrangian given by
\be
{\cal L}_{\rm brane} =\sqrt{-g} \Big[\alpha_1 K+ \alpha_2 R + \alpha_3 {\cal K}_{\rm GB} + \alpha_4 R_{\rm GB} \Big],
\label{potential3}
\ee
where $R$ is the Ricci scalar, $R_{\rm GB}$ is the Gauss-Bonnet term,
$R_{\rm GB} = R^2 - 4 R_{\mu \nu}^2 + R_{\mu \nu \alpha \beta}^2$,
$K$ is trace of the extrinsic curvature, and the ${\cal K}_{\rm GB}$ is the boundary term associated with the five dimensional
  Gauss-Bonnet term: $
{\cal K}_{\rm GB} = K^3-3 K K_{\mu \nu}^2+ 2 K_{\mu \nu}^3$.
Using the expression for the intrinsic curvature in terms of the extrinsic curvature,
 Eq.~(\ref{gauss}), the Lagrangian (\ref{potential3}) can be written as
 \be
\mathcal{L}_{\rm brane}\,
=\,-\sum_{n=1}^4 \beta_n\det_n(K),
\label{4Dbrane}
\ee
where $\beta_n= -n! \alpha_n$, thus it has {\it  exactly} the same structure  as Eq.~(\ref{lagpi}). We then denote the position modulus of the probe 3-brane as $\pi$. The induced metric on the brane is determined by $\pi$ as
\be
g_{\mu \nu} = \eta_{\mu \nu} + \partial_{\mu} \pi \partial_{\nu} \pi,
\label{induce}
\ee
and the extrinsic curvature is given by
\be
K_{\mu \nu} =  \gamma \partial_{\mu} \partial_{\nu} \pi, \;\;
\gamma=\frac{1}{\sqrt{1 + (\partial \pi)^2}}.
\ee
If we take a limit $\partial \pi \ll 1$, i.e. $\gamma \rightarrow 1$,  we find that the extrinsic curvature is simply  $K_{\mu \nu} = \Pi_{\mu \nu}$. Then the Lagrangian (\ref{4Dbrane}) reduces to (\ref{lagpi}).
This suggests that there may be a higher-dimensional interpretation behind the
Lagrangian (\ref{lagpi}).  Although we find these arguments very compelling, so
far we have not been able to pursue these
connections further, and for this reason we will
not develop them in this work. But
we should note that what we discussed follows the same construction as in
 the so called DBI Galileon  \cite{deRham:2010eu}.
  If we do not take the limit $\gamma \rightarrow 1$, the Lagrangian (\ref{4Dbrane}) becomes non-trivial, and it reproduces the Galileon terms. Since the four
  dimensional Gauss-Bonnet
   piece is a total derivative, there is no contribution from this term even away from
   the $\gamma \to 1$ limit.

\smallskip

After this digression to five dimensions,
we would like to return to Lagrangian (\ref{lagpi}), and discuss how to
render it  fully covariant.
 To do so, we need to understand how to go back from the field $\pi$ to the
original  St\"uckelberg fields
 $\phi^\mu$, in order
 to  restore the dependence of the vector mode and the full metric. Here we follow the
 approach discussed in Ref~\cite{claudia}. If only the scalar mode
$\pi$ is considered, then we can solve for $\Pi_{\mu\nu}$ in terms of $\phi^\mu$,
using (\ref{H}) and (\ref{Hscalar}). 
 The result is a second order algebraic equation for $\Pi_{\mu\nu}$, with solution $\Pi^{\ \nu}_{\mu} \,=\,\Lambda_3\left[
 \delta^{\ \nu}_{\mu} - \left(\sqrt{(\partial\phi)\eta (\partial\phi)^{T}}\right) _\mu ^{\ \nu}\right]$, where  $[(\partial\phi)\eta (\partial\phi)^T]_\mu^{\ \nu}\equiv\partial_\mu \phi^\alpha \partial^\nu \phi^\beta\eta_{\alpha \beta}$, and just to remind the reader $\Lambda_3^3=m^2M_{pl}$. The previous expression is written in terms of the square root of a matrix, formally understood as $\sqrt{M}_\mu^{\ \alpha}\sqrt{M}_\alpha^{\ \nu}=M_\mu^{\ \nu}$.

   \smallskip

We now go beyond the pure scalar sector case, and {\it define}
\be\label{K}
{\mK}^{\ \nu}_{\mu} \equiv \delta^{\ \nu}_{\mu} - \left(\sqrt{(\partial\phi)\eta (\partial\phi)^{T}}\right) _\mu ^{\ \nu}
=\delta^{\ \nu}_{\mu} - \left(\sqrt{\mathbb{I} - g^{-1} H }\right)^{\ \nu}_{\mu}
=\delta_\mu^{\ \nu}-\left(\sqrt{ g^{-1} [\eta +Q] }\right)^{\ \nu}_{\mu}.
\ee
 Notice that the previous quantity contains also contributions from vector
 and tensor degrees of freedom.
 On the other hand,
by construction ${\mK}^{\mu}_{\nu}$ becomes $\Pi^{\mu}_{\nu}\,/\Lambda_3$
 when only the scalar mode is considered.  The full non-linear Lagrangian for massive gravity is
then  constructed by substituting $\Pi$ by $\mK$ in (\ref{lagpi}). Namely
 \be
{\cal L}_{\mK} = -
\Big[
\alpha_1 \det_{1} ({\cal K}) +2\alpha_2 \det_{2} ({\cal K}) + 6\alpha_3 \det_{3} ({\cal K}) + 24\alpha_4 \det_{4} ({\cal K})
\Big]\label{potentialU}
\ee
where $\alpha_n=-n!\beta_n$, and the determinants $\det_n(\mK)$ are defined in Eqs.~(\ref{dets}) with $A=\mK$. The second term, for positive $\alpha_2$, reduces to the Fierz-Pauli term (\ref{lagFP}) when expanding the Lagrangian in terms of $H_{\mu \nu}$ around the Minkowski metric. Therefore, since we would like to have the Fierz-Pauli as the first correction to Einstein's gravity at leading order, we are not be interested
on the contributions from the first term, $\det_1 \mK$. Then, from now on we set $\alpha_1=0$.
  As a result, a family of non-linear massive gravity Lagrangians can be written as
\be\label{genlag}
{\cal L} = \frac{M_{Pl}^2}{2}\,\sqrt{-g}\left( R -2\Lambda - m^2{\cal U}\right),
\ee
where ${\cal U}={\cal L}_\mK$ with $\alpha_1=0$ and $\alpha_2=1$. These Lagrangians are parametrised by $m$, $\alpha_3$ and $\alpha_4$; moreover
we added a bare cosmological constant $\Lambda$.

In this family of Lagrangians, there is a special choice of parameters:
 $\alpha_3=-1/3$ and $\alpha_4=1/12$. It corresponds to the choice $z=-1$ in the expansion (\ref{detA}) with $A=\mK$. The Lagrangian is
\be\label{nambueq}
{\cal L}_{NG}=2 m^2\sqrt{-g} (\det(\mathbb{I} -\mK) - \tr \mK ) =2 m^2
\Big(
\sqrt{-\det(\partial_\mu\phi^\alpha\partial_\nu\phi^\beta\eta_{\alpha\beta})}
- \tr \mK \Big),
\ee
where the first term is the Nambu-Goto type of action for a bosonic 3-brane \cite{witten}. However, we will discuss in what follows that this particular choice, even though has a striking physical interpretation, does not allow to recover the GR solutions via the Vainshtein mechanism.

\smallskip

The $\alpha_2$ term in the the potential (\ref{potentialU}) was first suggested in \cite{claudia} as non-linear completion of FP theory. It
was shown that this term reduces to a particular choice of Galileon terms in the decoupling limit, without any coupling between the scalar mode and the massless graviton. Formally, the decoupling limit corresponds to a limit in which the scale $\Lambda_3\,=\,m^2 M_{pl}$ is kept fixed, while sending $M_{pl}\to \infty$
and $m\to 0$.
 Once $\alpha_3$ and $\alpha_4$ are included,
    the picture changes in an interesting way,
     and  couplings between $\pi$ and the massless
  graviton appear, even in the decoupling limit.
  These couplings
     have important theoretical and observational consequences; as
  we will discuss
in detail in the next Section.
   Interestingly,  these mixing terms are finite
in number, and do not spoil the fundamental
property that the equations of motion for $\pi$ are second order \cite{deRham:2010ik}.
  It is still an open question whether this remains true away from the decoupling limit,
  ensuring the absence of ghost degrees of freedom.
   A full Hamiltonian analysis should be carried out to set a final word on the subject. However, there are hints that the theory is ghost-free perturbatively, and for the particular choice of $\alpha_3=\alpha_4=0$ this has been shown  up to and including quartic order in perturbations \cite{claudia}.

\smallskip

We have now the necessary ingredients
 to discuss the second problem addressed in the Introduction,
 namely the vDVZ discontinuity.

 \section{Vainshtein mechanism at work}\label{sect-vain}

 In \cite{us}, we showed that Vainshtein mechanism applies for
  Lagrangians as (\ref{genlag}), setting to zero the coefficients $\alpha_3$ and
   $\alpha_4$, and the bare cosmological constant. (See Ref.~\cite{Damour:2002gp} for spherical symmetric solutions in the FP theory). Here we extend the analysis
   to arbitrary coefficients. We determine stringent
   constraints  on the parameter space of these
   theories, in order to ensure that the Vainshtein mechanism works.
   As we are going to discuss,
   our results  find a natural  interpretation
     in terms of the dynamics of
   helicity-zero mode, in the decoupling limit.

   \smallskip

In order to discuss solutions associated with
 Lagrangian (\ref{genlag}), it is convenient to express $\mathcal K$,
 given in Eq.~(\ref{K}),
  in terms
of matrices as
\be
\mK = {\mathbb I}-\sqrt{ \mM},
\ee
where ${\mathbb I}$ denotes the identity matrix and $\mM = g^{-1} \left[\eta+\mathcal Q\right]$. The task is to calculate
the trace of $\mM^n$.  Given that $\mM$ is a square matrix, it is diagonalizable, and can be expressed as
$\mM \,=\,{\cal U} \,D \,{\cal U}^{-1}$, for some invertible matrix ${\cal U} $, where $D$ is a diagonal matrix containing the eigenvalues of $\mM $. We shall
call the eigenvalues $\lambda_1,\dots,\lambda_4$.
Then, since $  \mM^n\,=\,{\cal U} \,D^n \,{\cal U}^{-1}$,
the traces in the formulae above can be expressed in terms of eigenvalues
\be
\tr{\mM^n}=\sum_i \lambda_i^n,
\ee
and the traces of $\mK^n$ result
\bea
\tr \mK &=& 4 - \tr \sqrt{\mM}, \nonumber\\
\tr \mK^2 &=& 4 -2 \tr \sqrt{\mM} + \tr \mM, \nonumber\\
\tr \mK^3 &=& 4 - 3 \tr \sqrt{\mM} + 3 \tr \mM - \tr \mM^{3/2}, \nonumber\\
\tr \mK^4 &=& 4 -4 \tr \sqrt{\mM} + 6 \tr \mM -4 \tr \mM^{3/2} + \tr\mM^2.
\eea
 Using the formulae for expressing
  the determinants in terms of traces (\ref{dets}), we can easily construct the potential.

We now discuss the
conditions    to recover GR results  in the small graviton mass limit,
  within a certain radius from a mass source.
   In particular, we are interested to determine under which circumstances the Vainshtein
   mechanism applies.
   In order to do so, we study spherically symmetric perturbations around flat space, expressed
  in spherical coordinates as
   $ds^2 = -dt^2 + dr^2 + r^2 d \Omega^2$,
     with $d \Omega^2 = d \theta^2 + \sin^2 \theta d \phi^2$.
     We start our discussion using the unitary gauge, $\pi^\mu=x^\mu-\phi^\mu=0$. Consider the following Ansatz for the metric
\be
ds^2 = - N(r)^2 dt^2 + F(r)^{-1} dr^2 + r^2 H(r)^{-2} d \Omega^2,
\label{diagonal}
\ee
that reduces the potential in (\ref{genlag}) to  
\bea\label{potential1}
 \sqrt{-g} \;{\cal U}
= - && \!\!\!\!\!\!\!\!\!
  \frac{r^2 \sin(\theta)}{\sqrt{F} H^2}\bigg\{ 2\left[\sqrt{F} \left((2 H-3) N+1\right)+H^2 N+H (2-6 N)+6 N-3\right] \nonumber\\ &&\hskip1.8cm
  - 6 \alpha_3 (H-1) \left[\sqrt{F} ((H-3) N+2)-2 H N+H+4 N-3\right]
   \nonumber
  \\
  &&\hskip1.8 cm
  -24 \alpha_4 (1-\sqrt{F}) (1-H)^2 (1-N)\bigg\}.
\eea
Notice that in GR one can set $H(r)=1$ by a coordinate transformation, but this is not possible here, since we have already chosen a gauge. The field equations are obtained by varying the action (\ref{genlag}) with respect to $N,\ H$ and $F$.
  The $N$-equation is the Hamiltonian constraint, so it only depends on $F$ and $H$. Since the equation for $H$ is quite complicated, we instead consider  a combination of the three equations which  gives $\nabla^\mu G_{\mu\nu} = 0$, where $G_{\mu\nu}$ is the Einstein tensor; it corresponds to the Bianchi identity. Therefore, we work with the Hamiltonian constraint, the Bianchi identity and the equation for $F$.
  The corresponding expressions are
   lengthy, so we relegate them to Appendix A.

    First, let us study solutions in the weak field limit, by expanding $N,\ F$ and $H$ as
\be
N = 1 + n, \;\; F = 1 + f, \;\; H= 1+ h,
\ee
and truncating the field equations to first order in these perturbations.
As we will see in what follows, this linearisation
procedure is not completely consistent for all values of the radial coordinate $r$, and
we will need to improve it.
In order to analyse the system,
it is convenient to introduce a new radial coordinate
\be
\rho = \frac{r}{H(r)}\,,
\label{chcoord}
\ee so that the linearised metric is expressed as
\be
ds^2 = - (1 + 2  n) dt^2 + (1 - \tilde{f}) d\rho^2
+ \rho^2 d \Omega^2,
\ee
where $\tilde{f} =  f - 2  h - 2 \rho  h'$ and a prime denotes a derivative with respect to $\rho$. As discussed above, one should be careful with this change of coordinates, since, after fixing a gauge, a change of frame in the metric modifies the St\"uckelberg field $\pi^\mu$ as well. However, for the moment, let us focus on the
 change of the
 metric part; later we will discuss what happens to $\pi^\mu$. At linear order, the equations for the functions $n(\rho)$, $\tilde{f}(\rho)$ and $h(\rho)$ in the new variable $\rho$ are
\bea\label{Neq}
0&=& \left(m^2\rho^2+2\right)\tilde{f}+2 \rho \left(\tilde{f}'+m^2 \rho^2
h'+3 m^2 \rho h \right), \\
0&=&  m^2 \rho^2 (n-2 h) -2\rho  n'-\tilde{f}, \label{feq}
\\
0&=&  \tilde{f} +\rho n'\label{const}.
\eea
In this linear expansion, the solutions for $n$ and $\tilde{f}$ are
\bea
2 n &=& - \frac{8 G M}{3 \rho} e^{- m \rho}, \nonumber\\
\tilde{f} &=& -\frac{4 G M}{3 \rho} (1 + m \rho) e^{- m \rho},
\label{linsol}
\eea
where we fix the integration constant so that $M$ is the mass of a point particle at the origin, and $8 \pi G = M_{pl}^{-2}$. These solutions exhibit the vDVZ discontinuity,  since the post-Newtonian parameter $\gamma=f/2n$ is $\gamma=1/2(1+m\rho)$, which in the massless limit reduces to $\gamma=1/2$, in
 disagreement with GR, and
 with  Solar system observations ($\gamma = 1$ in GR, while
  observations provide $1-\gamma\simeq 10^{-5}$  \cite{Will:2005va}).

However, in order to understand what really happens in this limit, we must
 also analyse  the behaviour of $h$ as $m\rightarrow 0$.
 For doing this, we consider scales below the Compton wavelength $m \rho \ll 1$, and at the same time ignore higher order terms in $G M$. Under these approximations, the equations of motion can still be truncated to linear order in $\tilde{f}$ and $n$, but since $h$ is not necessarily small, we have to
  keep all non-linear terms in $h$.  The resulting equations are then (see Appendix A for their derivation) 
  \bea\label{Neqfull}
&&\hspace{-1cm} 0= 2 \tilde f+2\rho \tilde f' +m^2\rho^2\Big(\left[ 1-2(3\alpha_3+1)h+3 (\alpha_3 +4\alpha_4)h^2\right] \left[(2+ \tilde f) \rho h'+(1+h) \tilde f\right]
\nonumber \\  &&\hspace{3cm} +6 h [1-(3\alpha_3+1)h+(\alpha_3 +4\alpha_4)h^2] \Big) \;,
\\ \label{feqfull}
&&\hspace{-1cm} 0=-\tilde f-2\rho n' +m^2\rho^2\Big(n -2 [1+n+(3 \alpha_3+1) n]h+[(3 \alpha_3+1)(n+1)+ 3(\alpha_3+4 \alpha_4)n]h^2 \Big) \;,
\\ \label{constfull}
&&\hspace{-1cm} 0=\rho n' \left[-1+2(3\alpha_3+1)h-3 (\alpha_3+4 \alpha_4)h^2 \right]- \tilde f [1-(3\alpha_3+1)h] \; .
\eea

  \smallskip

We start with the $N$-equation ($\ref{Neqfull}$). Since it only depends on $\tilde{f}$ and $h$, one can solve for $\tilde{f}$ in terms of $h$, including all non-linear terms in $h$. The solution, dropping higher order terms proportional to $m^4$, $(GM)^2$ and $m^2GM$, is 
\be\label{solf}
\tilde{f} = - 2 \frac{G M}{\rho} - (m \rho)^2 \left[
h - (1+ 3\alpha_3)h^2+(\alpha_3+4\alpha_4)h^3\right]\;.
\ee
Then we take the second equation (\ref{feqfull}),  obtained by varying the action with respect to $F$, and use the solution (\ref{solf}) for $\tilde{f}$. We find an
 expression for $n$ as a non-linear function of $h$. It
 turns out  to be  simpler to work with $n'$,  given by 
\be\label{soln}
2 \rho\,  n' = \frac{2 G M}{\rho} - (m \rho)^2 \left[
h - (\alpha_3 +4\alpha_4) h^3\right]\;,
\ee
where again we have dropped terms with $m^4$, $(GM)^2$ and $m^2GM$.
Finally, the constraint equation (\ref{constfull}) gives an equation for $h$,
 after substituting the solutions for $n$ and $\tilde{f}$ given
 respectively
 by (\ref{soln}) and (\ref{solf}). We should stress that this equation for $h$ is
  {\it  exact},
so there are no higher order corrections. It is given by  
\bea
\frac{G M}{\rho} \left[1 - 3(\alpha_3+4\alpha_4)h^2\right]&=& - (m \rho)^2
\Big\{\frac{3}{2} h -  3 (1 + 3\alpha_3)h^2  +
\left[(1 + 3\alpha_3)^2 + 2(\alpha_3+4\alpha_4)\right]h^3 \nonumber\\
&&\hskip1.8cm - \frac{3}{2}(\alpha_3+4\alpha_4)^2h^5 \Big\}\;.
\label{solh}
\eea
If we linearise the equations (\ref{solf}), (\ref{soln}) and (\ref{solh}) with respect to $h$, we recover the solution (\ref{linsol}); on the other hand, below the so-called Vainshtein radius
\be \rho_V \,=\, \left(\frac{G M }{m^{2}}\right)^{1/3}\;,\ee $h$ becomes larger than one.
Therefore, in this regime, we have to include higher order contributions
due to $h$ to equations  (\ref{solf}) and (\ref{soln}).
 There are three qualitatively
 different cases,
  depending on the values of the
    parameters $\alpha_3$ and $\alpha_4$:
\begin{itemize}
\item {\it Case with $\alpha_3=-1/3, \alpha_4= 1/12$}.
This is a special situation, since all higher order contributions in $h$ vanish from equations (\ref{solf})-(\ref{solh}). Therefore, there is no Vainshtein effect and the solutions to the equations are those given in (\ref{linsol}) for $\rho<1/m$. This model, corresponding
 to the bosonic 3-brane Lagrangian (\ref{nambueq}),
is ruled out by Solar system observations.

\item {\it Case with $\alpha_3=-4\alpha_4\neq -1/3$}.
For $\rho \ll \rho_V$ we can solve for $h$ from the last equation (\ref{solh})
keeping only the highest order terms in $h$. Then, the solution for $h$ is 
\be
h=-\frac{1}{(1 -12\alpha_4)^{2/3}}\frac{\rho_V}{\rho},
\ee
which implies $|h| \gg 1$ for $\rho \ll \rho_V$, as expected. We can then use this solution and equations (\ref{solf}) and (\ref{soln}) to give the expressions for $n$ and $\tilde{f}$ within the Vainshtein radius, namely  
\bea
2 n &=& -\frac{ 2 G M}{\rho}\, \left[1 - \frac{1}{2(1 - 12 \alpha_4)^{2/3}} \left(\frac{\rho}{\rho_V}
\right)^2 \right]\;,\nonumber \\
\tilde{f} &=& - \frac{ 2 G M}{\rho} \left[1 - \frac{1}{2(1 - 12\alpha_4)^{1/3}} \left(\frac{\rho}{\rho_V}
\right) \right]\;.\label{solfbv}
\eea
Therefore, the corrections to GR solutions are indeed small for $\rho$ smaller than the Vainshtein radius $\rho_V$.

\item {\it Case with $\alpha_3\neq-4\alpha_4$}. This case can be divided in two: $\alpha_3+4\alpha_4>0$ and $\alpha_3+4\alpha_4<0$. 
 The latter is the most intriguing case, and we focus on it. In the limit $\rho \ll \rho_V$, the solution for $h$ is given by
\be
h = -\left(\frac{2}{\alpha_3+4\alpha_4}\right)^{1/3}\frac{\rho_V}{\rho}
-\frac{\left[ 2\left(1+3 \a_3\right)^2+3  \left(\a_3+4 \a_4\right) \right]}{9
\,\left[2 \left(\a_3+4 \a_4\right)^5\right]^\frac13}
\,\frac{\rho}{\rho_V}\;\;\;,
\ee
so that $|h| \gg 1$. Notice that, in solving
Eq.~(\ref{solh}), we include also the next-to-leading order, that
 results to be linear in $\rho$. 
   It turns out that this expression for
  $h$ provides a correction of the same order  as $GM/\rho$ in the $n$ and $\tilde{f}$ equations.
 Indeed, plugging this expression in the equations for $n$ and $\tilde f$,
 one gets
\bea
2 n &=& {\cal O} \Big(\Big(\frac{\rho}{\rho_V}\Big)^2 \Big)\frac{GM}{\rho}
\;,\nonumber \\
\tilde{f} &=& {\cal O} \Big(\frac{\rho}{\rho_V} \Big)  \frac{GM}{\rho}
\;.\label{solfbv2}
\eea

Surprisingly, the contribution from the scalar mode $h$ exactly cancels the usual $1/\rho$ potential at  leading order. So, gravity becomes {\it weaker} approaching
the source, for distances  smaller than Vainshtein radius, and larger
than Schwarzschild radius. This implies that the strong coupling of the scalar graviton not only shields interactions of the scalar mode $h$, but also those of
the massless graviton.
As we will see later while analysing the decoupling limit, this is related to a coupling between the scalar mode and graviton that cannot be removed by a local field transformation  if $\alpha_3 \neq -4 \alpha_4$ \cite{deRham:2010ik}. In this case, local tests of gravity would
  completely rule out the predictions
   of the theory at leading order in $h_{\mu\nu}$. However, when $\alpha_3+4\alpha_4>0$, apart from the previous solution, a second solution exists in the limit $\rho \ll \rho_V$ \cite{david}, where $h=1/\sqrt{3(\alpha_3+4\alpha_4)}+\mathcal{O}(\rho^3)$. In this case, the metric fields $n$ and $f$ present small corrections to the GR solution within the Vainshtein radius, and the solution cannot be ruled out by Solar system tests. 

   In order to fully understand the system with $\alpha_3\neq-4\alpha_4$, it is imperative to study how the two possible solutions extend beyond the Vainstein radius, both in vacuum and in the presence of matter. This is beyond the scope of the present analysis and will be studied in a future publication \cite{future1}. 


\end{itemize}

\noindent
To summarize, only the choices $\alpha_3 +4 \alpha_4> 0$ and $\alpha_3 =-4 \alpha_4\neq\,-1/3$
allow to recover standard GR in the weak field limit, 
below the Vainshtein radius $\rho_V$. This fact imposes a stringent constraint in parameter space for the theory described by Lagrangian (\ref{genlag}).

\smallskip

Now, we would like to go back to the issue of the coordinate transformation introduced in Eq.~(\ref{chcoord}),
 and discuss another way to interpret the
previous results.
 As  mentioned earlier, a coordinate transformation introduces a change in $\pi$ of the form (\ref{pitransf}). When changing $r/H=\rho$, we excite the radial component of the St\"uckelberg field, as $\pi^{\rho} = - \rho \,h$. Thus the strong coupling nature of $h$ is encoded in $\pi^{\rho}\,=\,\eta^{\rho \rho}\,\partial_\rho \pi/\Lambda_3^3$, 
 when working with the coordinate $\rho$. As
 a result,
  the non-linear analysis previously done is more transparent in the decoupling limit,
  as developed in  \cite{deRham:2010ik}.
As previously mentioned, this limit is achieved by taking $m \to 0$ and $M_{pl} \to \infty$, while keeping $\Lambda_3=(m^{2}M_{pl})^{1/3}$ fixed. This implies that, when
substituting $H_{\mu\nu}$ back into the full Lagrangian (\ref{genlag}), 
 we do  not consider the vector mode, and expand the potential
 at linear order in $h_{\mu \nu}$. 
   The resulting expression describes  the theory in the decoupling limit, and  contains
    the quadratic Hilbert-Einstein piece,  total derivatives in $\pi$ given by (\ref{lagpi}), and finally  mixing terms between $h_{\mu \nu}$ and $\pi$. The
     mixing terms
     are
     
\be
h^{\mu\nu}\sum_i X^{(i)}_{\mu\nu}, \qquad \qquad \mathrm{with}\qquad \sum_iX^{(i)}_{\mu\nu}=\frac{\partial {\cal L}_\mK}{\partial h^{\mu\nu}}\Big|_{h_{\mu\nu}=0},
\ee
and each $X^{(i)}$
is of order
${\cal O}(\Pi^i)$. There are only three  mixing contributions,
 which can be absorbed (except for $X^{(3)}$) in the remaining
 terms by the non-linear field redefinition  
\be
h_{\mu \nu} = \hat{h}_{\mu \nu} +\frac{\pi}{M_{pl}} \eta_{\mu \nu}
- \frac{1 +3\alpha_3}{\Lambda_3^3 M_{pl}} \partial_{\mu} \pi \partial_{\nu} \pi.
\label{metric}
\ee
After this field redefinition, the resulting Lagrangian is ($[ \Pi^n ]\,\equiv
\tr{\Pi^n}$) \cite{deRham:2010ik}

\bea\label{decouplingaction}
{\cal L} &=& {\cal L}_{\rm GR} (\hat{h}_{\mu \nu})
+ \frac{3}{2} \pi \Box \pi - \frac{3(1 +3\alpha_3)}{2 \Lambda_3^3}
(\partial \pi)^2 \Box \pi +
\frac{(\partial \pi)^2}{2 \Lambda_3^6}\Big[
(1 +3\alpha_3)^2+2(\alpha_3+4\alpha_4)\Big]\,
([\Pi^2] - [\Pi]^2) \nonumber\\
&&- \frac{5}{4 \Lambda_3^9} (1+ 3\alpha_3)(\alpha_3+4\alpha_4)(\partial \pi)^2
([\Pi]^3 - 3 [\Pi] [\Pi^2]+2 [\Pi^3])+\frac{M_{pl}}{\Lambda_3^6}\hat{h}^{\mu\nu}X^{(3)}_{\mu\nu},
\eea
where ${\cal L}_{\rm GR}$ is the quadratic Einstein-Hilbert action for $\hat{h}_{\mu \nu}$, and 
\bea
X^{(3)}_{\mu\nu}&=& - \frac{1}{2}(\alpha_3 + 4\alpha_4)\Big\{
6\Pi_{\mu\nu}^{3}-6(\tr\Pi)\Pi_{\mu\nu}^2 + 3\left[
(\tr\Pi)^2
-(\tr\Pi^2)\right]\Pi_{\mu\nu} \nonumber\\
&&\hskip2.5cm- \left[(\tr\Pi)^3 - 3(\tr\Pi^2)(\tr\Pi)+2(\tr\Pi^3)\right]\eta_{\mu\nu}\Big\}\;.
\label{fidex3}
\eea
Interestingly, the Lagrangian  obtained so far  in  the decoupling limit allows us to re-interpret  the constraints
 on $\alpha_3$ and $\alpha_4$
obtained earlier in this Section. One can observe that the remaining direct coupling between
 the scalar $\pi$ and graviton $\hat h_{\mu\nu}$ either vanishes or is negative for the two cases which successfully
 implement Vainshtein mechanism.  


Observe also that the kinetic terms for $\pi$ in (\ref{decouplingaction}) are precisely the Galileon terms, which give rise to  second order differential equations for $\pi$ \cite{Nicolis:2008in}. The non-linear structure of these terms involving $\pi$ is
essential to recover GR in some range of scales distances; for the choice $\alpha_3
=-4\alpha_4=-1/3$,  
we obtain an Einstein frame Lagrangian for Brans-Dicke gravity with a vanishing Brans-Dicke parameter and no potential.
However, this particular choice is not compatible
with observations, as we have already mentioned during our general analysis, in the
first part of this Section.

\smallskip

We would like now to further analyse the equations of motion for this system
 in  the decoupling limit.
 The Lagrangian (\ref{decouplingaction}) is exact in the decoupling limit, as there are no higher order terms besides those shown, so the linearised Einstein equation for $\hat{h}_{\mu \nu}$ is 
 \be
 \delta G_{\mu \nu}(\hat{h}) - \frac{1}{M_{pl} \Lambda_3^6} X^{(3)}_{\mu \nu}=0.
 \ee
 If we assume spherical symmetry, then it can shown that $X^{(3)}_{\mu \nu}$ is simply  
 \be
 X^{(3)}_{tt} = - \frac{\alpha_3 + 4 \alpha_4}{\rho^2} (\pi'^3)', \quad
 X^{(3)}_{\rho\rho} = 0.
 \ee
Therefore, the solutions for the linearised metric, $\hat{h}_{tt}=- 2 \hat{n}$
and $\hat{h}_{\rho\rho} = - \hat{f}$, reduce to
\bea
\hat{f} &=& -\frac{2 G M}{\rho} + \frac{\rho^2}{\Lambda_3^6 M_{pl}} (\alpha_3 + 4 \alpha_4)
\Big(\frac{\pi'}{\rho}\Big)^3, \nonumber\\
2 \rho \hat{n}' &=& - \hat{f}.
\label{hatmetric}
\eea
On the other hand, the equation of motion for $\pi$ derived
from the action (\ref{decouplingaction}), is given by (see \cite{Nicolis:2008in})
\bea
3 \left(\frac{\pi'}{\rho}\right)
&& \!\!\!\!\!\!\!
+ \frac{6}{\Lambda_3^3}(1 + 3\alpha_3)\left(\frac{\pi'}{\rho}\right)^2
+  \frac{2}{\Lambda_3^6}\Big[
(1 + 3\alpha_3)^2+2(\alpha_3+4\alpha_4)\Big] \left(\frac{\pi'}{\rho}\right)^3
\nonumber\\
&&
+ \frac{6 M_{pl} (\alpha_3 + 4 \alpha_4)}{ \rho^2  \Lambda_3^6} ( \rho \hat{n}')
 \left(\frac{\pi'}{\rho}\right)^2
 = \frac{M}{4 \pi \rho^3 M_{pl}}
\;,\nonumber\\
\label{pieq0}
\eea
where the integration constant is again chosen so that $M$ is a mass of a particle at the origin. Using the relation between $\pi$ and $h$, $h = -\pi'/\left(m^2 M_{pl} \rho\right)$, it is simple to check that
the solutions for $\tilde{f}$, $n$ and $h$ given by Eqs.~(\ref{solf})-(\ref{solh}) agree with
the expressions in
 Eqs.~(\ref{metric}), (\ref{hatmetric}) and (\ref{pieq0}). This confirms  that the results
 obtained earlier in this Section are in perfect agreement with what is found
 in terms of the dynamics
 of the scalar field $\pi$, in the decoupling limit.

\smallskip

In summary, the Vainshtein mechanism applies for $\alpha_3 +4 \alpha_4> 0$ and $\alpha_3 =-4 \alpha_4\neq\,-1/3$. Only for these choices the weak field GR results are fully recovered at distances smaller than the Vainshtein radius. If these solutions match the asymptotic configurations (\ref{linsol}) beyond the Vainshtein radius, then there would be three phases: on the largest scales
   beyond Compton wavelength,
   $ m^{-1} \ll \rho$, the gravitational interactions decay exponentially due to the graviton mass, see Eq.~(\ref{linsol}).
In the intermediate region $\rho_V < \rho < m^{-1}$, we obtain the $1/r$ gravitational potential but Newton's constant is rescaled, $G \to 4 G/3$. Moreover, the post-Newtonian parameter reduces to $\gamma=1/2$ in the $m\rightarrow 0$ limit, instead of $\gamma=1$ of GR, showing the vDVZ discontinuity. Finally, below the Vainshtein radius $\rho<\rho_V$, the GR solution is recovered due to the strong coupling of the $\pi$ mode (see, for example, Eq.~(\ref{solfbv}) for the case $\alpha_3 =-4 \alpha_4\neq\,-1/3$); in this regime the theory can then be rendered compatible with observations.

The solution discussed here provides  a testing arena
 for studying the Boulware-Deser ghost. Instead of expanding the action in $H_{\mu \nu}$ around Minkowski spacetime up to higher orders in perturbations, we have the
 possibility to
   study {\it linear} perturbations around this non-perturbative solution, using the complete potential in  Eq.~(\ref{potentialU}). In order to obtain the full
    non-linear solution, matching the three phases we have described,
     a numerical approach is necessary.
      In the next section, we consider a different family of vacuum
      solutions for  this theory,
       which can be obtained analytically, and can lead to interesting
       candidates for realistic backgrounds.

\section{Exact solutions}\label{sec-exact}

As we learned in the previous section, an essential property of this theory
of massive gravity
is the strong coupling phenomenon occurring
 in the proximity of a source. This allows, for certain regions of parameter
space, to recover linearised General Relativity at sufficiently small distances by
means of the Vainshtein  mechanism.
    This behaviour, accompanied
 by  the fact that Birkhoff theorem
  does not apply in this context,    suggests that exact solutions for this theory,
  even imposing spherical symmetry, might be very different from the GR
  ones.

  In this section, we will exhibit new
  spherically symmetric exact solutions in the vacuum for  massive gravity,
  that generalize the ones of \cite{Salam:1976as} and \cite{us}.
  In an appropriate gauge, the solutions are asymptotically de Sitter or Anti-de Sitter,
  depending on the choice of parameters.

  While in \cite{us} we focused  on the case $\alpha_3=\alpha_4=0$, we now generalize
 the analysis to arbitrary couplings in the Lagrangian (\ref{potentialU}).  We adopt the unitary gauge
  and allow for arbitrary couplings
  $\alpha_i$, $i=2,..,4$ (as explained earlier, we set to zero
   the coefficient
    $\alpha_1$). We start with the following form
 for the metric (for convenience, we
 implement slightly different conventions with respect to the previous
 section)
 \be\label{metrans}
 d s^2\,=\,- C(r) \,d t^2+2 D(r)\,d t dr +A(r)\, d r^2+ B(r)\,d \Omega^2
 \ee
  so that, even though the spacetime is spherically symmetric, the metric contains a cross term
  $d t dr$.
  We choose the following Ansatz for the metric
  functions \cite{Salam:1976as, us},

  \bea
  B(r)&=& b_0\, r^2\,,\nonumber
  \\
  C(r)&=& c_0 +\frac{c_1}{r}+c_2 \,r^2\,,\nonumber\\
  A(r)+C(r)&=& Q_0\,,\nonumber\\
  D^2(r)+A(r) C(r)&=&\Delta_0 \label{ansmetrcomp},
  \eea
  and use the equations of motion to fix the
  constant parameters $b_0, c_0, c_1, c_2, Q_0, \Delta_0$.
   Einstein equations read
   \be
   G_{\mu\nu}\,=\, 8 \pi G \,T_{\mu \nu}
   \ee
   with
    energy momentum tensor
   $T_{\mu\nu}\,=\,- \frac{1}{\sqrt{-g}}\frac{\delta {\cal L}_K}{\delta g^{\mu\nu}}$, and
   ${\cal L}_K$ given in Eq.~(\ref{potentialU}).

  In General Relativity, diffeomorphism invariance allows one to choose the function $B(r)$ to be
  $B(r) \,=\,r^2$ , so that $b_0=1$. In
   this theory of massive gravity, after having fixed
  the gauge, this choice is no longer possible and the equations
  of motion determine $b_0$.
    In order to do this, one observes
  that the metric Ansatz (\ref{metrans})
  leads to the following identity between
  components of the
  Einstein tensor:  $C(r) G_{rr}+A(r) G_{tt}\,=\,0$.
  This combination on the energy momentum tensor provides the following
  value for
 $b_0$,
 \be\label{solb0}
b_0 \,=\,
\left(
 \frac{1 + 6 \a_3 + 12 \a_4 \pm\sqrt{1 + 3 \a_3 + 9 \a_3^2 - 12
\a_4}}{3 (1+3 \a_3 +4 \a_4)}
\right)^2.
\ee
The upper branch
generalizes the result of \cite{us}, while the lower
branch is specifically associated
 with theories in which $\alpha_3$ and/or $\alpha_4$ are non-vanishing.
    After Plugging the metric components
(\ref{ansmetrcomp}) in the remaining Einstein equations,
one can find the values for the other parameters. The corresponding
general
 expressions are quite lengthy, and for this reason we relegate them
 to Appendix \ref{AppB}.
 As a concrete, simple
 example, in the main text we
  work out the special case
    we focussed on the previous section, $\a_3=-4 \,\a_4$.
In this case, a solution is given by the following
 values  for the parameters

 \bea
 {b_0} &=&\frac{4}{9}\left(
 \frac{1-12 \a_4}{1-8 \a_4}
 \right)^2\nonumber,\\ 
 c_0&=&\frac{\Delta_0}{b_0}\nonumber,\\
 c_2&=&\frac{m^2\,\Delta_0}{4\left(12 \a_4-1\right)}
 \nonumber,\\
 Q_0&=&
 \frac{
 16 (1 - 12 \a_4)^4 + 81 (1 - 8 \a_4)^4 \Delta_0}{36 \left[
 1 + 4 \a_4 (-5 + 24 \a_4)\right]^2}.
 \eea 
The previous solution is valid for $\alpha_4$ in the ranges
$\alpha_4< 1/12$ and $\alpha_4 > 1/8$. Notice that the case $\alpha_4
= 1/12$ corresponds exactly to the Lagrangian (\ref{nambueq}), discussed in 
Section \ref{covsect}. 
  We find that
  $c_1$ and $\Delta_0$ are arbitrary; this vacuum solution is then
  characterized by two integration constants.
The resulting metric coefficients
 can be rewritten in the following, easier-to-handle form:
\bea
A(r)&=&\frac94 \,\Delta_0\,
\left(\frac{1-8 \a_4}{1-12 \a_4}\right)^2\,\left[ p(r) +\gamma+1\right]
\hskip0.5cm,\hskip0.5cm B(r)\,=\,\frac{4}{9}\,
\left(\frac{1-12 \a_4}{1-8 \a_4}\right)^2\,r^2\nonumber\\\label{solrew}
C(r)&=&\frac94 \,\Delta_0\,
\left(\frac{1-8 \a_4}{1-12 \a_4}\right)^2\,\left[1- p(r) \right]
\hskip0.5cm,\hskip0.5cm \\D(r)&=&
\frac{9 \Delta_0}{4} \,\left(\frac{1-8\a_4}{1-12\a_4}\right)^2\,
\sqrt{p(r) \left(p(r)+\gamma\right)}\nonumber
\eea
with ($\mu=-c_1/c_0$)
\be\label{defpb}
p(r)\,\equiv\,\frac{\mu}{r}+\frac{\left(1-12 \a_4\right)\,m^2\,r^2}{9\,(1-8 \a_4)^2}
\,\hskip0.8cm,\hskip0.8cm \gamma\,\equiv\,\frac{16}{81 \Delta_0}\,
 \left(
\frac{1-12 \a_4}{1-8 \a_4}\right)^4
-1.\ee

In order to have a consistent solution, we must demand
 that the argument of the square root appearing in the expression
 for $D(r)$, Eq.~(\ref{solrew}),  is positive. A  sufficient condition to ensure
 this is that $\mu \ge 0$, and
 \be
 0\, <\,
  \sqrt{\Delta_0}\,<\,\frac{4}{9}\,
  \left(
\frac{1-12 \a_4}{1-8 \a_4}\right)^2\,.
 \ee

\smallskip

The metric then could be rewritten in a more transparent diagonal
form, by means
of a coordinate transformation. However, a coordinate transformation
of the time coordinate is not permitted,  since until this
point we have adopted the
unitary gauge. Therefore, we  now renounce to this gauge choice,
and allow for a non-zero vector
$\pi^\mu$ of the form $\pi^\mu\,=\,\left( \pi_0(r),0 ,0 ,0\right)$. One finds that
then the metric can be rewritten in a diagonal form, as
\be\label{newfmetr}
d s^2\,=\,-C(r) d t^2+\tilde{A}(r) d r^2+B(r)\,d \Omega^2,
\ee
while the equations of motion for the fields involved are solved by
\be
\tilde{A}(r)\,=\,\frac{4}{9}\,
\left(\frac{1-12 \a_4}{1-8 \a_4}\right)^2\frac{1}{1-p(r)}
\hskip0.8cm,\hskip0.8cm \pi_0'(r)\,=\,-\frac{\sqrt{p(r) (p(r)+\gamma)}}{1-p(r)}, 
\ee
with $C(r)$ and $B(r)$ being the same as in Eq.~(\ref{solrew}).
If one makes  a time
rescaling $$t\to \frac{ 4
\left(1-12 \a_4\right)^2 }{
9 \Delta_0^{1/2}
\left(
1-8 \a_4\right)^2}\,t\; ,
$$
 the resulting metric has then a manifestly de Sitter-Schwarzschild,
or Anti-de Sitter-Schwarzschild form. This depends on whether  $\alpha_4$
is smaller or larger than $1/12$, as can be seen inspecting
the function $p(r)$ in Eq.~(\ref{defpb}).
 On the other hand, we should point out that this time-rescaling
cannot be performed, without further introducing a time dependent
contribution to $\pi_0$. As expected, the metric in Eq.~(\ref{newfmetr})
can also be obtained by making the following transformation
of the time coordinate $d \tilde t\,=\,d t +\pi_0' dr$ to the original
 metric
(\ref{metrans}). This produces a non-zero time component for
$\pi^\mu$, that does not vanish even in the limit $m\to 0$.

\smallskip

To summarize so far, we found vacuum solutions in this theory that
are  asymptotically de Sitter or Anti-de Sitter, depending
on the choice of the parameters (Another family of  solutions
with similar behaviour, but obtained for different choices of parameters,
have been recently discussed in  \cite{Nieuwenhuizen:2011sq}).
Let us point out that it is also possible to include
a bare cosmological constant term $  \sqrt{-g}\,
\Lambda$ to the Lagrangian (\ref{genlag}). Our solutions
to the Einstein equations, with our metric Ansatz, remain
formally identical.
The only difference is that the function $p(r)$ in Eq.~(\ref{defpb})
 becomes
 \be
 p(r)\,=\, \frac{\mu}{r}+\frac{\left(1-12 \a_4\right)}{9\,(1-8 \a_4)^2}
 \,\left[m^2+\frac43 \left(1-12 \a_4\right)\Lambda
 \right]\,r^2.
 \ee
Notice that the additional integration constant $\Delta_0$
can not be used to 'compensate' the contribution
of the  bare cosmological
constant $\Lambda$ via a self-tuning mechanism, since $\Delta_0$ does not explicitly appear in the
previous formula.   For asymptotically de Sitter solutions, $\alpha_4<1/12$,
 choosing
$\mu=0$,
the metric can also be written in a time dependent form, at the price
of switching on additional components of $\pi^\mu$, as mentioned earlier. After dubbing  
$$\tilde m^2\equiv
\frac{1}{(1-12 \a_4)}
 \,\left[m^2+\frac43 \left(1-12 \a_4\right)\Lambda
 \right],
$$
we can make the following coordinate transformation
$t\,=\,F_t(\tau, \rho)$ and $r\,=\,F_t(\tau, \rho)$ with

\bea
F_t(\tau, \rho)&=&\frac{4}{3 \,\Delta_0^{1/2} \,\tilde{m}}\,
\Big(\frac{1-12 \alpha_4}{1- 8 \alpha_4} \Big)
\,{\rm arctanh}\,\left(
\frac{\sinh{\left(\frac{\tilde m \tau}{2}\right)}+\frac{\tilde{m}^2 \rho^2}{8} e^{
\tilde{m} \tau/2}}{\cosh{\left(\frac{\tilde m \tau}{2}\right)}-\frac{\tilde{m}^2 \rho^2}{8} e^{
\tilde{m} \tau/2}}
\right),
\\
F_r(\tau, \rho)&=&\frac32 \Big(\frac{1-8 \alpha_4}{1-12 \alpha_4}\Big)\rho \, e^{
\tilde{m} \tau/2}.
\eea
The metric becomes that of flat slicing of de Sitter
\be
  d s^2\,=\,
- d \tau^2+ e^{{\tilde m} \tau}\,\left( d \rho^2+\rho^2  d \Omega^2\right),
\ee
where the Hubble parameter is given by
\be
H\,=\,\frac{\tilde m}{2}\,=\,\frac{1}{2 (1-12 \a_4)^{\frac12}}
\,\left[m^2+\frac43 \left(1-12 \a_4\right)\Lambda
\right]^\frac12.
\ee
The St\"uckelberg fields $\pi^\mu$ are now given by
$\pi^\mu\,=\,\left( \pi^\tau(\tau, \rho), \, \pi^\rho(\tau, \rho),\,0,\,0 \right)$,
with $\pi^\tau\,=\,\pi_0+\tau-F_t$, $\pi^\rho\,=\rho-\,F_r$.
Interestingly, the value of the Hubble parameter
  is ruled by the mass of the graviton; in the case
 of vanishing bare cosmological constant,
 we have a self-accelerating solution, in which
   the smallness of the
 observed cosmological constant can be associated with the smallness
 of the graviton mass.

\smallskip
This self-accelerating solution, which reduces to that found perturbatively in Ref.~\cite{deRham:2010tw},
 appears as  an ideal background
 to explain present-day acceleration. 
Notice that this configuration is remarkably similar to that in the DGP braneworld model \cite{Deffayet:2000uy}, though there are important differences. In order to study the viability of our non-perturbative solution, 
 it is necessary to study the behaviour of fluctuations, to confirm that there is no ghost. On the other hand, in the DGP model, the self-accelerating solution suffers from a ghost instability \cite{Luty:2003vm, Nicolis:2004qq, Koyama:2005tx}, which is related to the ghost in the FP theory on a de Sitter background. 


\section{Discussion}\label{sect-discus}

In this work, we investigated the consequences of strong coupling effects for a theory of non-linear massive gravity, developed by de Rham, Gabadadze and collaborators. We first re-expressed the complete potential for this theory in a particularly compact and easy-to-handle form. Among other things, this way of writing the potential suggested
intriguing relations with a higher dimensional set-up, that might offer new perspectives. Moreover, we showed that, for certain parameter choices, the potential for massive gravity coincides with the action describing non-perturbative brane objects, independently
supporting connections with a higher dimensional framework.

We then studied the conditions to implement
the Vainshtein mechanism in this context. The theory is able to reproduce
the behaviour of linearised General Relativity below a certain scale, but only in a specific region of parameter space. This result provides stringent constraints on this non-linear massive gravity models. Moreover, we were able to physically re-interpret our findings in the decoupling
limit, in terms of an effective theory with Galileon interactions for the scalar-graviton. We showed that the  condition to successfully implement Vainshtein mechanism is associated with the sign of a direct coupling between the massless graviton and scalar degree of freedom which cannot be removed by a local field transformation.

We also presented new exact solutions in the vacuum for this theory that asymptote either
de Sitter or anti de Sitter space, depending on the choice of the parameters.  
Asymptotically de Sitter configurations can be expressed in an explicit time-dependent form, providing an interesting background for the observed Universe,
where the rate of the accelerated expansion of the Universe is set by the graviton mass.
A small graviton mass, as required by the solar system constraints on deviations from standard General Relativity, is then in agreement with the observed value of the cosmological constant. On the other hand, asymptotically anti de Sitter configurations may have interesting AdS/CFT applications.

\smallskip
Our results naturally lead to various important questions to be future examined.
It would be interesting to put in a firmer basis the connection between this theory of non-linear massive gravity and the higher dimensional set-up described in Section \ref{covsect}. Exploiting this relation might also shed light on the absence of ghost excitations at all orders in perturbations about Minkowski. In Section \ref{sect-vain}, we found the intriguing result that, in some region of parameter space, the coupling between the scalar degrees of freedom and graviton shields the interactions of linearised graviton so that gravity appears to become {\it weaker} as the source is approached. To understand in full detail the system in this case, it is however necessary to analyse the dynamics of higher order metric fluctuations and their couplings with the scalar sector. Although it seems unlikely that strong coupling dynamics of gravitational modes behave such to mimic General Relativity in relevant regimes (by producing a sort of Vainshtein effect at higher order in perturbations), we cannot exclude this possibility for all ranges of parameters. This paper provides all the necessary equations to study this deep issue, away from the decoupling limit. It would also be interesting to study in more detail the exact vacuum configurations discussed in
Section \ref{sec-exact}, in particular to understand whether de Sitter solution controlled by the graviton mass can be rendered stable under perturbations. If so, it can be considered as a serious candidate for the observed universe where the present day acceleration of the
Universe is only due to gravitational degrees of freedom.

\subsection*{\it Note Added}

While this work was in the last stages of preparation, two papers \cite{Hassan:2011vm, Nieuwenhuizen:2011sq}
appeared, with some overlap with our Section \ref{covsect} with respect to the
formulation of the potential for non-linear massive gravity in terms
of determinants.

\subsection*{Acknowledgements}

We would like to thank David Pirtskhalava for discussions. KK and GN are supported by the ERC. KK is also supported by STFC grant ST/H002774/1, while GT is supported by STFC Advanced Fellowship ST/H005498/1. KK acknowledges support from the Leverhulme trust.

\begin{appendix}

\section{Equations of motion for spherically symmetric and asymptotically flat backgrounds}
Here we give details of the equations used in Section 2 to describe spherically symmetric solutions with the theory defined by (\ref{genlag}). Using the Ansatz
\be
ds^2 = - N(r)^2 dt^2 + F(r)^{-1} dr^2 + r^2 H(r)^{-2} d \Omega^2
\ee
in the action (\ref{genlag}), we obtain the potential
\bea
\sqrt{-g} \;{\cal U}
= - && \hspace{-.6cm}
  \frac{r^2\sin(\theta)}{\sqrt{F} H^2}\bigg\{ 2\left[\sqrt{F} \left((2 H-3) N+1\right)+H^2 N+H (2-6 N)+6 N-3\right] \nonumber\\
 && \hskip1.8cm
  - 6 \alpha_3 (H-1) \left[\sqrt{F} ((H-3) N+2)-2 H N+H+4 N-3\right]
   \nonumber
  \\
  &&\hskip1.8 cm
  -24 \alpha_4 (1-\sqrt{F}) (1-H)^2 (1-N)\bigg\}.
\eea
Varying it with respect to $N$ gives the equation of motion
\bea\label{Neqn}
0=&&\hspace{-.6cm}
3\alpha_3 m^2 r^2 (1-H) H^2 \left[\sqrt{F} (H-3)-2 H+4\right]- 12 \alpha_4 m^2 r^2 \left(\sqrt{F}-1\right) (1-H)^2 H^2
   \nonumber
\\&& \hspace{-.3cm} -H^2 \left[r \left(\dot{F}-6 m^2 r\right)+3 m^2 r^2\sqrt{F} \nonumber
   +F\right]+r H \left(r \dot{F} \dot{H}+2 F \left(r \ddot{H}+3
   \dot{H}\right)\right)\\ && \hspace{-.3cm} -5  r^2 F\left(\dot{H}\right)^2+2 m^2 r^2
   \left(\sqrt{F}-3\right) H^3 +H^4 \left(m^2
   r^2+1\right),
\eea
where `dot` denotes derivatives with respect to $r$. Varying with respect to $F$ gives
\bea\label{Feqn}
0=&&\hspace{-.6cm} -3\alpha_3 m^2 r^2 (1-H) H^2  \left[H (2 N-1)-4 N+3\right] \nonumber
-12 \alpha_4 m^2 r^2 (1-H)^2 H^2 (1-N)  \\ && \hspace{-.3cm} +2 r F H \dot{H} \left(r \dot{N}+N\right)-r^2 F N \left(\dot{H}\right)^2 +H^2
   \left[N \left(6 m^2 r^2-F\right)-r \left(2 F \dot{N}+3 m^2
   r\right)\right]\nonumber\\&& \hspace{-.3cm} +H^4 N \left(m^2 r^2+1\right)+2 m^2 r^2 H^3 (1-3
   N),
\eea
and finally by varying with respect to $H$, one gets
\bea
0=&& \hspace{-.6cm}  6 \alpha_3 m^2 r H^2  \left[\sqrt{F} (H (2 N-1)-3 N+2)+H (2-3 N)+4
   N-3\right]   \nonumber \\ && \hspace{-.3cm} - H^2 \left[r \dot{F} \dot{N}+N \dot{F}+2 F \left(r \ddot{N}+\dot{N}\right)+2 m^2 r \sqrt{F}
   (3 N-1) -12 m^2 r N +6 m^2 r\right] \nonumber \\ && \hspace{-.3cm} +H \left(rN  \dot{F}\dot{H}+2 F
   \left(rN  \ddot{H}+r \dot{H} \dot{N}+2 N \dot{H}\right)\right)-4r F N
   \left(\dot{H}\right)^2 \nonumber \\ &&  \hspace{-.3cm}
   +2 m^2 rH^3  \left[\left(\sqrt{F}-3\right)
   N+1\right] +24 \alpha_4 m^2 r \left(\sqrt{F}-1\right) (H-1) H^2  (N-1).
\eea
Instead of using the last equation, it is simpler to use a constraint based on the (second) Bianchi identity. This can be achieved by taking a combination of the previous field equations which leads to $\nabla^\mu G_{\mu\nu}=0$ for the Einstein piece, where $G_{\mu\nu}$ is the Einstein tensor. The constraint constructed in this way is
\bea\label{const1}
0=\frac{1}{rHN}\bigg\{&& \hspace{-.6cm}  -3 \alpha_3 \Big[\sqrt{F} \Big[H \left(N (4 r \dot{H}+6)
-2 r \dot{H}+3 r \dot{N}-4\right)+2 r (2-3 N)  \dot{H} \nonumber \\ && \hspace{-.3cm} +r H^3 \dot{N}  +H^2 (-4 r \dot{N}-4
   N+2)\Big]+2 H^2 (H (2 N-1)-3 N+2)\Big] \nonumber \\ && \hspace{-.3cm}
+12 \alpha_4 (1-H) \left[\sqrt{F} \left[2 r (N-1)  \dot{H}+r H^2 \dot{N}-H (r \dot{N}+2 N-2)\right]
+2 H^2 (N-1)\right]\nonumber \\ && \hspace{-.3cm} +\sqrt{F} \left[-H \left[2 N (r \dot{H}+3)+3 r \dot{N}-2\right]+2 (3 N-1) r \dot{H}+2 H^2 (r \dot{N}+N)\right]\nonumber \\ && \hspace{-.3cm} -2 H^2 [(H-3) N+1]\bigg\}.
\eea
We would like to study perturbations about flat space, hence
\be
N = 1 + n, \;\; F = 1 + f, \;\; H= 1+ h,
\ee
and study linear perturbations. However, $n$ and $f$ are small, like in GR, but $h$ could be large since this is where the scalar graviton has an influence. Therefore, we will keep higher orders in $h$ and truncate the equations to first order in $n$ and $f$. It is more convenient to introduce a new radial coordinate $\rho = r/H$, so that the linearised metric is expressed as
\be
ds^2 = - (1 + 2  \tilde{n}) dt^2 + (1 - \tilde{f}) d\rho^2
+ \rho^2 d \Omega^2.
\ee
The change of coordinates fixes $\tilde{f}$ in terms of $f$, as $
\tilde{f}(\rho) = f(r(\rho)) \left\{ \partial_\rho \left[
\rho H(\rho)\right]\right\}^2$,
while we have the freedom to choose $\tilde{n}$ and $\tilde{h}$ in terms of $n$ and $h$. For simplicity, we pick $\tilde{n}(\rho)=n(r(\rho))$ and $\tilde{h}(\rho)=h(r(\rho))$. Therefore, we can drop the tildes of $n$ and $h$ from now on, and $\tilde{f}$ simplifies to
\be
1-\tilde{f}(\rho)=[1-f(r(\rho))][1+h(\rho)+\rho h']^2,
\ee
where a prime denotes a derivative with respect to $\rho$. Then, equation (\ref{Neqn}), in the new variable $\rho$ and to leading order in $\tilde{f}(\rho)$ but keeping all orders in $h(\rho)$, reads  
\bea
0= 2 \tilde f+2\rho \tilde f' +m^2\rho^2\Big(&& \hspace{-.6cm}\left[ 1-2(3\alpha_3+1)h+3 (\alpha_3 +4\alpha_4)h^2\right] \left[(2+\tilde f) \rho h'+(1+h) \tilde f\right]
\nonumber \\ && \hspace{-.3cm} +6 h [1-(3\alpha_3+1)h+(\alpha_3 +4\alpha_4)h^2] \Big) .
\eea
Equivalently for equation (\ref{Feqn}), one gets
\be
0=-\tilde f-2\rho n' +m^2\rho^2\Big(n -2 [1+n+(3 \alpha_3+1) n]h+[(3 \alpha_3+1)(n+1)+ 3(\alpha_3+4 \alpha_4)n]h^2 \Big).
\ee
Finally, for the constraint equation (\ref{const1}), the expression to first order in $\tilde{f}$ and $n$, but to all orders in $h$ is
\be
0=\rho n' \left[-1+2(3\alpha_3+1)h-3 (\alpha_3+4 \alpha_4)h^2 \right]-\tilde f [1-(3\alpha_3+1)h].
\ee
These last three equations are the ones used in Section 2 to derived the Vainshtein mechanism.

\section{General exact solution}\label{AppB}

In order to construct exact solutions to the Lagrangian (\ref{genlag}), we use the following Ansatz
\be
 d s^2\,=\,- C(r) \,d t^2+2 D(r)\,d t dr +A(r)\, d r^2+ B(r)\,d \Omega^2,
\ee
with
\bea
B(r)&=& b_0\, r^2\,,\nonumber
  \\
C(r)&=& c_0 +\frac{c_1}{r}+c_2 \,r^2\,,\nonumber\\
A(r)+C(r)&=& Q_0\,,\nonumber\\
D^2(r)+A(r) C(r)&=&\Delta_0 .
\eea
The constant parameters $b_0, c_0, c_1, c_2, Q_0, \Delta_0$, can then be fixed using Einstein equations
\be
   G_{\mu\nu}\,=\, 8 \pi G \,T_{\mu \nu}
\ee
with  $T_{\mu\nu}\,=\,- \frac{1}{\sqrt{-g}}\frac{\delta {\cal L}_K}{\delta g^{\mu\nu}}$, and ${\cal L}_K$ given in Eq.~ (\ref{potentialU}). The combination $C(r) G_{rr}+A(r) G_{tt}\,=\,0$ fixes uniquely $b_0$ to be
\be\label{boapp}
b_0 \,=\,
\left(
 \frac{1 + 6 \a_3 + 12 \a_4 +\Gamma_\pm}{3 (1+3 \a_3 +4 \a_4)}
\right)^2,
\ee
where
\be\label{gamma}
\Gamma_\pm\equiv \pm\sqrt{1 + 3 \a_3 + 9 \a_3^2 - 12 \a_4}.
\ee
By requiring that the $1/r^3$ term in the $G_{tt}$ equation vanishes, we obtain
the condition
\be
c_0=\frac{\Delta_0}{b_0},
\ee
and using the rest of Einstein's equations leads to a solution for the remaining coefficients. These are
\bea
&& \hspace{-.5cm}c_2=\frac{\Delta_0 m^2}{9 b_0(1+3 \alpha_3+4 \alpha_4)^2}\Big[1-2\Gamma_\pm+4 \alpha_4 (2 \Gamma_\pm-7)
+2\alpha_3 (1-18   \alpha_4-2 \Gamma_\pm)
+\alpha_3^2 (15-6 \Gamma_\pm) +18 \alpha_3^3\Big]
 \nonumber\\
&& \hspace{-.5cm}Q_0= \frac{1}{81 b_0(1+3 \alpha_3+4 \alpha_4)^4}\Big[ 8 + 8 \Gamma_\pm + 81 \Delta_0 12 +\alpha_3 \big[10 + 81 \Delta_0 + 2^6 3^3 \alpha_4^3 (2 + 3 \Delta_0) + 9 \Gamma_\pm \nonumber\\ && \hspace{4cm} +
    216 \alpha_4^2 (1 + 18 \Delta_0 + 4 \Gamma_\pm) + 6 \alpha_4 (17 + 162 \Delta_0 + 22 \Gamma_\pm)\big]
\Gamma_\pm
\nonumber\\ && \hspace{4cm} +
 27 \alpha_3^2 \big[27 + 162 \Delta_0 + 288 \alpha_4^2 (5 + 9 \Delta_0) + 20 \Gamma_\pm +
    8 \alpha_4 (29 + 162 \Delta_0 + 26 \Gamma_\pm)\big]
\nonumber\\ && \hspace{4cm}
+ 2^7 3^3 \alpha_4^3 (-1 + 6 \Delta_0 + 2 \Gamma_\pm)
+2^8 3^4\alpha_4^4 (1 +\Delta_0) + 81 \alpha_3^4 (41 + 81 \Delta_0)
\nonumber\\ && \hspace{4cm}
+ 54 \alpha_3^3 (41 + 162 \Delta_0 + 24 \alpha_4 (14 + 27 \Delta_0) + 20 \Gamma_\pm)
\Gamma_\pm \nonumber\\ && \hspace{4cm}
+ 144 \alpha_4^2 (1 + 54 \Delta_0 + 8 \Gamma_\pm) \Gamma_\pm
+ 48 \alpha_4 (2 + 27 \Delta_0 + 3 \Gamma_\pm)
 \Big].
\eea
Notice that there are two branches, depending on the sign choice in $b_0$ (see equations (\ref{boapp}) and (\ref{gamma})).  
For $\alpha_3 = - 4 \alpha_4$, the upper (lower) branch solution must be taken for $\alpha_4<1/12$ $(\alpha_4 >1/12)$. The solution for $Q_0$ exists only if $Q_0$ satisfies the condition $4 \sqrt{\Delta_0} + 2 Q_0 >0$. Due to this condition, for $\alpha_3 = - 4 \alpha_4$, the solution is valid only for $\alpha_4$ in the ranges $\alpha_4 <1/12$ and $\alpha_4>1/8$.

\end{appendix}

\end{document}